# Magnetic structure driven by monoclinic distortions in the double perovskite $Sr_2YRuO_6$

P L Bernardo[1], L Ghivelder[1], H S Amorim[1], J J Neumeier[2], and S García[1]

[1] Instituto de Física, Universidade Federal do Rio de Janeiro, Rio de Janeiro, 21941-972, Brazil

[2] Department of Physics, Montana State University, Bozeman, MT 59717-3840, USA

E-mail: sgg@if.ufrj.br

**Abstract**

The monoclinic double-perovskite $Sr_2YRuO_6$ has recently gained a renewed interest in order to get a deeper insight into the exotic magnetic ground states associated with geometric frustration. Striking discrepancies between the spin order derived from the neutron diffraction refinements and the macroscopic magnetic and thermal responses is a major challenge that must be addressed. In this work, detailed neutron diffraction measurements as a function of temperature yield a completely different interpretation of the patterns. We show that at low temperatures a spin structure of the $K_2NiF_4$-type is an accessible configuration for the magnetic ground state. In the neighborhood of the magnetic transition this configuration evolves into a canted super-structure. The deduced temperature dependence of the canting angle exhibits two closely spaced peaks, which are in excellent agreement with the double peaks in the magnetic contribution to the specific heat and in the thermal expansion coefficient. We explain these features in terms of reorientation of the net ferromagnetic moment of the non-collinear spin state, due to the local breaking of the inversion symmetry promoted by the monoclinic distortions, with structural changes acting as the driving force.

## 1. Introduction

Geometrically frustrated magnetic materials, such as double perovskites with chemical formula $A_2BB'O_6$, with $B$ = Y, Ca, Na or Li and $B'$ = Ru, Mo, Os or Re, exhibit a variety of fascinating magnetic structures comprising valence-bond glass [1], spin ice [2], spin liquid [3], and spin glass states [4], in addition to long-range ferromagnetic (FM) or antiferromagnetic (AFM) order [5]. A key point for this complex behavior is that the magnetic ions residing on the B'-sites present a face-centered-cubic (*fcc*) symmetry, which is a three dimensional lattice based on edge sharing tetrahedra, the simplest case of geometric frustration [6]. However, the fact that several isostructural materials with the same value of spin $S$ of the magnetic ion show quite different magnetic responses reveals that a clear behavior pattern is not uniquely defined by frustration effects. Systematic studies of double perovskites with different $S$ values [7-9] reveal that although for $S = ½$ quantum fluctuations are expected to be enhanced [7], several compounds of the family exhibit long-range ferromagnetism or antiferromagnetism [5], with minimal evidence of geometric frustration or a valence-bond glass state [1]. For $S = 1$, a tuning of competitive superexchange interactions through slight structural changes leads either to low dimensional short-range spin order or to a spin-frozen ground state [7-8], while for $S = 3/2$ long-range order with very low variation in magnetic entropy emerges [9]. Therefore, a detailed understanding of these remarkable differences requires a careful correlation between the spin structure derived from neutron powder diffraction (NPD) refinements and results obtained from other techniques. A clear example comes from muon-spin rotation and NPD measurements in $La_2LiMoO_6$ [7], which yield contradictory results.

Despite the fundamental interest in the understanding of *fcc* antiferromagnets, in some cases, to the best of our knowledge, there is still a gap related to the validation of the NPD refinements by macroscopic measurements, and to determine whether or not geometric frustration has indeed a leading role in the establishment of the magnetic ground state. In strongly distorted perovskites crystallochemical details might introduce a significant contribution. The monoclinic double perovskite $Sr_2YRuO_6$, with a frustration index $f \sim 15$ [9], space group $P2_1/n$, with pentavalent $4d^3$ $Ru^{5+}$ ions in a high-spin state ($^4A_{2g}$) with $S = 3/2$, is a particularly suitable compound to address these topics. The *collinear* type-I AFM structure, derived from the only two NPD measurements reported [10,11], is inconsistent with magnetic measurements in single crystal [12] and polycrystalline samples [13-15], which conclusively show that this material is a weak-ferromagnet. A crucial open issue is the nature of a double peak in the specific heat at $T_{N1} = 28$ K and $T_{N2} = 24$ K, just below the magnetic transition temperature, $T_M = 32$ K. Recently, this feature was interpreted in terms of a two-step model in which full long-range type-I AFM order at $T_{N2}$ is preceded by two dimensional (2D) magnetic fluctuations sandwiched between alternate parallel layers exhibiting partial long-range order at $T_{N1}$ [11]. Following this approach, the ordered Ru moments in the layers are collinear for all temperatures and the coupling mechanism is completely based on geometric frustration considerations. Additional interest concerning $Sr_2YRuO_6$ is that this compound might host an exotic type of high-$T_C$ superconductivity, when Ru ions are partially replaced by Cu (< 15 at. %) [16]. It has been proposed that superconducting holes condensate in the nonmagnetic SrO layers [17]. This is still a controversial topic and a comprehensive knowledge of the magnetic ground state of the parent compound is of primary interest to understand how the superconducting order parameter would percolate through the pair-breaking effect of the Ru-magnetic layers.

In this work we present a detailed analysis of NPD patterns collected at various temperatures between 10 K and 300 K, with a fine temperature step of 1 K in the 20 K ≤ T ≤ 33 K range. The emergence and subsequent disappearance of a forbidden line for the type-I AFM order as the temperature is varied in the interval between $T_{N2}$ and $T_{N1}$ allows us to refine the patterns using an alternative model, leading to a different interpretation of the nature of the double peak in the specific heat in terms of a non-collinear spin state accompanied by a re-orientation of the associated net FM moment.

## 2. Experimental

The $Sr_2YRuO_6$ sample was synthesized by solid-state reaction. Stoichiometric amounts of high-purity $RuO_2$, $Y_2O_3$ and $SrCO_3$ were ground together thoroughly in an agate mortar, placed in an alumina crucible, and fired first at 934 °C for 24 hours in air. The resulting powder was ground again, pressed into pellets and heat treated in air at 1250 °C, with intermediate regrindings, and finally sintered for four days with a 1 °C/min cooling rate in an oxygen flow. NPD measurements were carried out on the time-of-flight diffractometer at the Powgen line of the Spallation Neutron Source research center at the Oak Ridge National Laboratory, USA, using a high-intensity pulsed beam. The sample was placed inside a vanadium can and held in a liquid helium cryostat. Neutron wave lengths λ = 1.066 Å and λ = 4.797 Å were used to scan a wide range of nuclear reflections and to measure in detail the $Q$ region for magnetic peaks, respectively.

## 3. Results and Discussion

The refinements of the NPD patterns were performed using both the propagation vectors $K_0 = (0,0,0)$ and $K_{1/2} = (½,½,0)$. These vectors were automatically selected by the *K-SEARCH* sub-program in FullProf and confirmed by the Le Bail method according to the magnetic reflections. Refinements using $K_0$ and assuming a *P2$_1$/n* space group for the magnetic reflections yield the reported type-I AFM order for T ≤ $T_{N2}$ and $T_{N1}$ < T < $T_M$, with good fit parameters, as shown in Table I for selected temperatures. However, it was not possible to refine the patterns for the interval $T_{N2}$ < T ≤ $T_{N1}$ using $K_0$.

When using $K_0$ and the *P2$_1$/n* space group for both nuclear and magnetic reflections only collinear spin configurations can be generated from the corresponding base vectors of the irreducible representations. We recall that although the monoclinic symmetry of the $Sr_2YRuO_6$ lattice is preserved on cooling, the bond distances and angles vary in a complex way [13], opening the possibility of non-collinear spin configurations as a consequence of unbalanced competitive FM and AFM interactions. These spin states would not necessarily exhibit the same periodicity of the lattice. Since $Sr_2YRuO_6$ is a weak ferromagnet, in an attempt to allow a possible canted spin structure to emerge from the refinements, we use the triclinic *P-1* space group for the magnetic reflections. In doing so, two Wyckoff positions for the Ru sites are considered and the corresponding Ru(1) and Ru(2) magnetic sub-lattices can be refined independently, with $\varphi_1$ and $\varphi_2$ being the azimuthal angles determining the respective orientation of the magnetic moments. Figure 1 shows the NPD pattern at T = 10 K as a function of $Q = (4\pi/\lambda)\sin\theta$, with $\lambda = 4.797$ Å (a) and $\lambda = 1.066$ Å (b). In addition to the nuclear Bragg peaks, the magnetic diffraction reflections were clearly identified, since they appear just below $T_M$. The intensity ratio of these magnetic reflections remains constant for T < $T_M$. The calculated pattern after magnetic and crystal structure refinements using $K_{1/2}$, the *P-1* space group for the magnetic reflections and keeping the *P2$_1$/n* space group for the nuclear ones, yield a very good agreement with the experimental data, as shown in Fig. 1. Weak reflections on both sides of the strong peak at ~ 2.2 Å$^{-1}$ correspond to a stray fraction of $SrY_2O_4$ (< 3%), that was identified and refined along with the $Sr_2YRuO_6$ profile. The fit parameters are quite similar to those obtained using $K_0$, as observed in Table I. However, the spin structure is of the $K_2NiF_4$-type [18], as shown in Fig. 2 for the elementary pseudocubic cell. In this configuration the orientation of the Ru moments located at the center of the faces is reverted in relation to the type-I AFM structure. A small canting angle, within the error of the measurement, between spins located in neighboring cells was observed. We will return to this point below. Therefore, the Ru moments in $Sr_2YRuO_6$ at T = 10 K have two accessible magnetic states, both with good fit statistics and meaningful physical parameters. For the $K_{1/2}$ case, the choice of the space group *P-1* for the magnetic reflections, keeping the *P2$_1$/n* space group for the nuclear ones, introduces a degree of freedom in the refinement that allows other spin configurations to be considered as possible ground states. Peak positions and relative intensities using $K_0$ and $K_{1/2}$ are not identical, but show only very slight differences within the corresponding refinement errors. Refinements at other temperatures in the intervals $T \leq T_{N2}$ and $T_{N1} < T < T_M$, using both $K_0$ and $K_{1/2}$ also show good matching.

The refinements for the interval $T_{N2} < T < T_{N1}$ are more complex and interesting. Figure 3(b) shows a sequence of NPD patterns at closely spaced temperatures within this region. Profiles in a wide temperature range are shown in Fig 3(a). This detailed temperature scanning allowed us to catch a remarkable feature: the emergence of a distinct magnetic reflection at T = 27 K, quite close to $T_{N1}$ = 28 K. This additional peak, located at Q = 0.86 Å$^{-1}$, increases its intensity on cooling to T = 26 K, diminishes for T = 25 K, and disappears for $T \leq T_{N2}$ = 24 K. It is a Gaussian symmetric reflection with line width given by instrumental resolution. Therefore, is not due to scattering from 2D-fluctuactions surviving below $T_N$. Its intensity is similar or even slightly higher as compared to other magnetic peaks detected for all temperatures below $T_M$. Most important, it was not possible to refine the NPD patterns for these three temperatures using $K_0$, while a good refinement was achieved using $K_{1/2}$. The fit parameters are presented in Table I. The resulting spin state is described by a canted magnetic superstructure along the *c*-axis, derived from the $K_2NiF_4$-type order previously obtained at low temperatures (see Fig. 2), and characterized by a canting angle $\eta = |\varphi_1-\varphi_2-180°|$, that measures the deviation from the antiferromagnetic orientation. The orientation of the Ru magnetic moments for this non-collinear state is shown in Fig. 4, which displays a top view from the *c*-axis of the pseudocubic cell. It is worth mentioning that the magnitude of the Ru magnetic moment shows a continuous and smooth decrease when crossing the $T_{N2}$ < T < $T_{N1}$ region as $T_M$ is approached on warming, as shown in the inset of Fig. 5. Therefore, the use of $K_{1/2}$ as propagation vector yields physically meaningful results for the whole temperature range below $T_M$.

A remarkable feature of the temperature dependence of the canting angle $\eta$, which reaches a high value of 85°±5°, is a well resolved double peak at temperatures quite close to $T_{N1}$ and $T_{N2}$, as shown in Fig. 5. The two maxima observed for the thermal expansion coefficient, $\alpha$, reported in Ref. [13] are also plotted for comparison. The excellent matching of the peaks unambiguously reveals that structural changes are involved in the re-arrangement of the spin configuration just below $T_M$, and clearly show the correspondence between the magnetic structure derived from the NPD refinements and the macroscopic properties. Additional support to the emergence of a canted spin state comes from the temperature dependence of the coercive field $H_C$ [14], which exhibits a maximum of ~2 kOe close to $T_{N1}$ followed by an abrupt decrease down to a few Oe near $T_{N2}$. We interpret this result as a consequence of reorientation of the net FM moment associated with the changes in the canted spin state detected by the variations in $\eta$. The collinear type I-AFM order cannot account for the $H_C$(T) behavior.

The role of the structural changes as the dominant mechanism in determining the magnetic ground state is supported by the fact that the two maxima in $C_{mag}$ do not change when a magnetic field H = 9 T is applied [14]. If, as proposed previously [11], the nature of these maxima was purely magnetic due to the transition from partial long-range spin order in alternate layers of the *fcc* Ru-sublattice at $T_{N1}$ to full long-range order at $T_{N2}$, it would be expected that a high field would leave a signature in this process. This is indeed the case of the double perovskite $Gd_2YRuO_6$, for which the double peak in $C_{mag}$ due to the magnetic coupling between the Gd- and Ru-sublattices shifts and smears by applying a magnetic field of a few Tesla [19]. On the other hand, if the proposed coupling mechanism is actually an inherent property of the Ru $fcc$ network, reflecting a general trend in $fcc$ antiferromagnets [11], the double perovskite $Ba_2YRuO_6$ must also present two maxima in the specific heat. However, only one λ-like sharp peak is observed [9,20]. In fact, $Ba_2YRuO_6$ would be the ideal compound to exhibit this two-step mechanism, since the strength of the next-nearest-neighbors Ru-O-O-Ru interactions, assumed to be the key point for alternate layered coupling, is favored by the cubic structure of this compound. In contrast, the monoclinic $Ca_2YRuO_6$ shows a broad peak for $C_{mag}$ with an unresolved structure that resembles the double peak observed for $Sr_2YRuO_6$ [22]. The smearing of the doublet is probably due to 44% of partial disorder between the Ca/Y ions, generating slightly different local neighborhoods for Ru ions [21]. Other Ru-based perovskites, such as monoclinic $Ca_2LaRuO_6$ and cubic $Ba_2LaRuO_6$, with similar frustration index $f \sim 18$, exhibit type I-AFM and type III-AFM order, respectively, confirming that the structural distortions are relevant [22].

## 4. Conclusions

In summary, careful refinements of neutron powder diffraction measurements show that the low temperature magnetic ground state in $Sr_2YRuO_6$ is of the $K_2NiF_4$-type, in contrast to previous studies. In the neighborhood of the magnetic transition temperature, the excellent matching between the double peaks in the canting angle of the Ru moments, the magnetic contribution to the specific heat, and the thermal expansion coefficient are interpreted in terms of a spin re-orientation of the net ferromagnetic moment associated with a non-collinear spin configuration. Comparison with the isospin compounds $Ba_2YRuO_6$ and $Ca_2YRuO_6$ supports that monoclinic distortions play a leading role in determining this canted structure, acting as the driving force of the magnetic response by unbalancing the fine equilibrium between ferromagnetic and antiferromagnetic interactions. Although geometric frustration is relevant, it cannot solely account for the observed features by simply considering the cancellation of the magnetic coupling between consecutive antiferromagnetic square layers. Therefore, the magnetic and thermal responses of $Sr_2YRuO_6$ do not reflect an inherent property of $fcc$ magnets.

# Acknowledgments

This research was supported by the Brazilian agencies FAPERJ, CNPq, and CAPES. We would like to thank Ashfia Huq, from Oak Ridge National Laboratory, for assistance with NPD measurements.

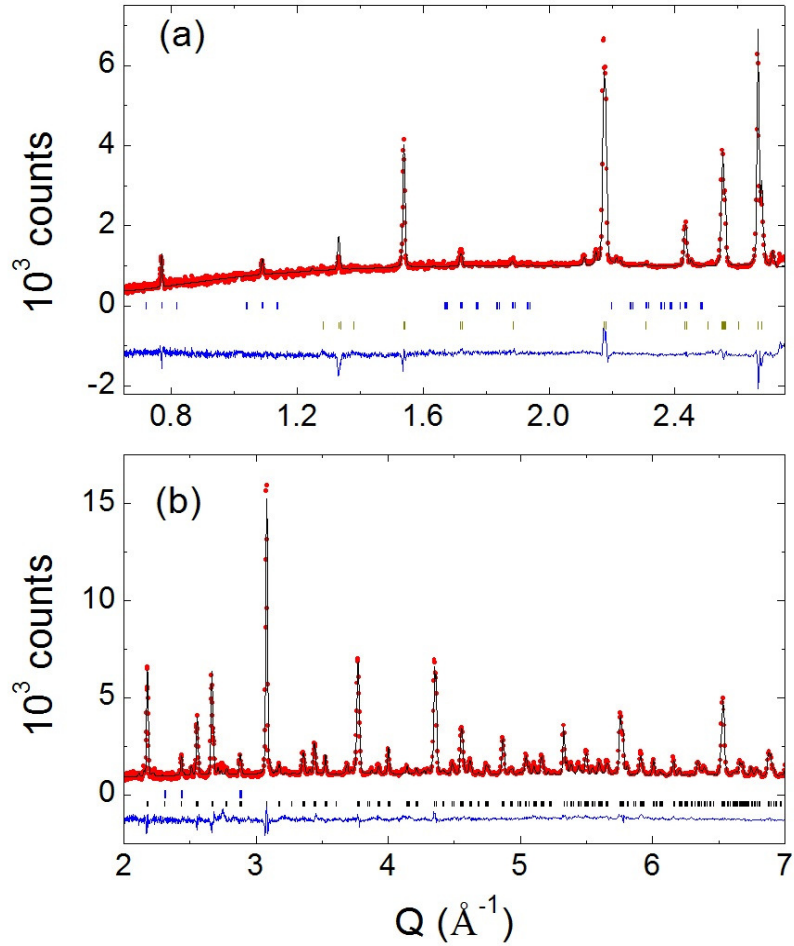

**Fig. 1** Neutron powder diffraction pattern of $Sr_2YRuO_6$ at 10 K for $\lambda$ = 4.797 Å (a) and $\lambda$ = 1.066 Å (b). The symbols and solid lines represent observed and calculated patterns, respectively. The difference curve is shown at the bottom. Vertical bars indicate the expected Bragg peak positions according to the magnetic (upper) and nuclear (lower) structure models described in the text. The Rietveld refinement was performed using the propagation vector $K_{1/2}$ = (½,½,0).

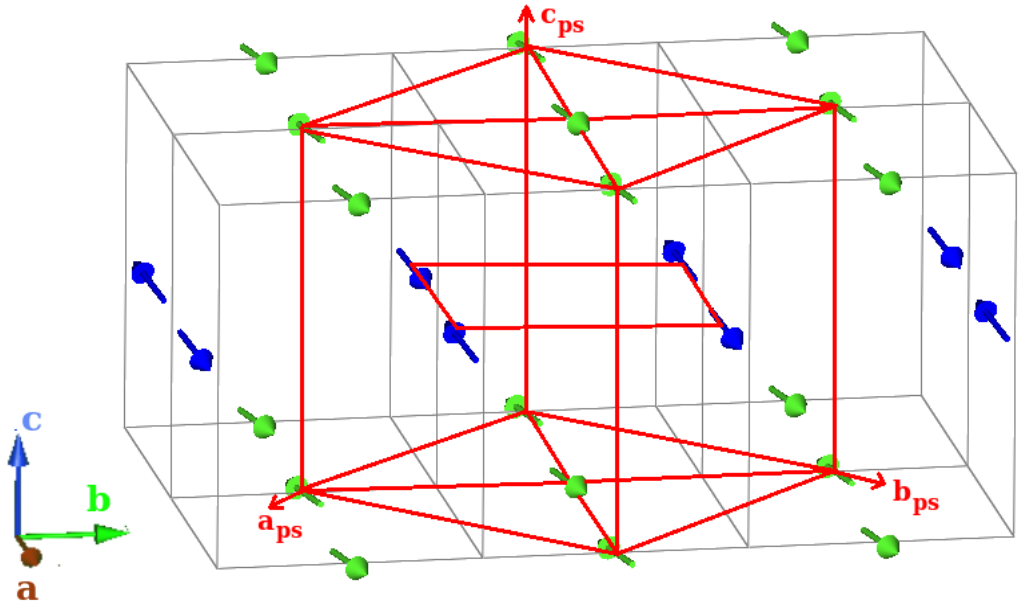

**Fig. 2** The orientation of the magnetic moments of the Ru ions for the K$_2$NiF$_4$-type spin structure obtained after Rietveld refinement of the diffraction patterns of Sr$_2$YRuO$_6$ at T = 10 K, using the propagation vector $K_{1/2}$ = (½,½,0). The pseudocubic (*ps*) unit cell is indicated and only the magnetic Ru ions are displayed. The orientation of the Ru moments located at the center of the faces is inverted in relation to the type-I AFM structure. The orientation of the monoclinic axes, *a*, *b*, and *c*, is indicated.

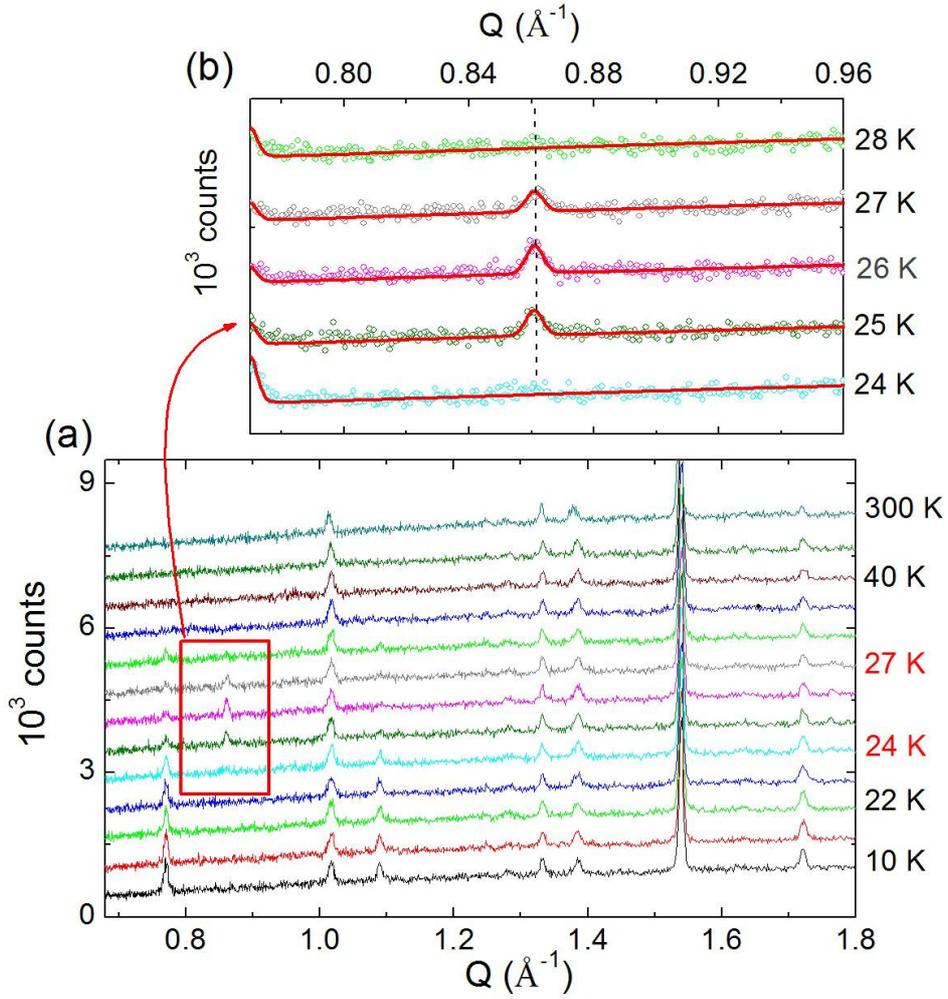

**Fig. 3** (a) Neutron powder diffraction patterns of $Sr_2YRuO_6$ for selected temperatures. The rectangle encloses a forbidden reflection for the type-I AFM structure at T = 25 K, 26 K and 27 K, located at $Q = 0.86$ Å$^{-1}$; (b): an enlarged section of the $Q$ region in the neighborhood of this magnetic reflection. The position of the peak is indicated by a dotted line. The continuous line is the calculated pattern for the full $Q$ range using the propagation vector $K_{1/2} = (½,½,0)$.

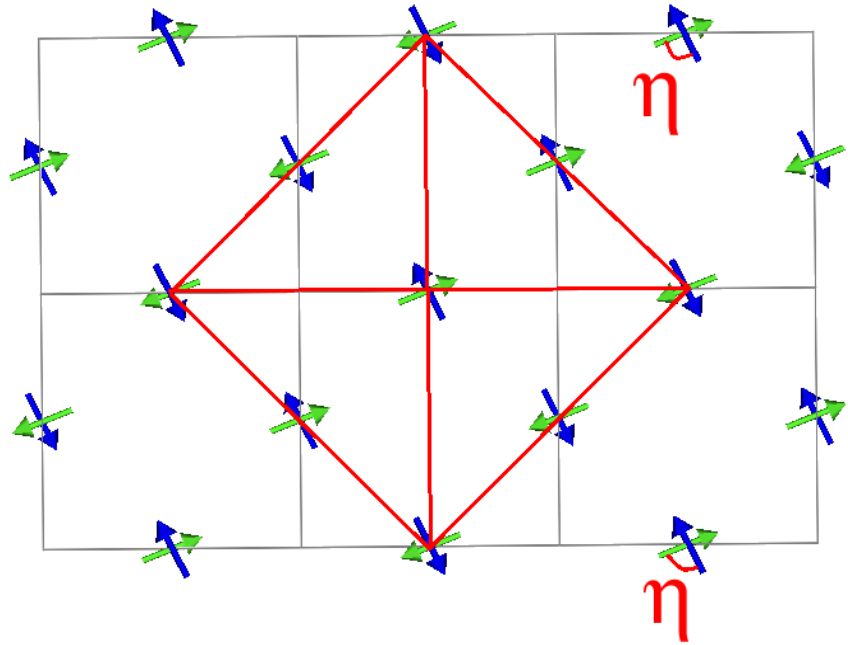

**Fig. 4** Top view from the *c*-axis of the pseudo cubic unit cell showing the orientation of the magnetic moments of the Ru ions for the canted spin state at T = 25 K using the propagation vector $K_{1/2}$ = (½,½,0). The non collinear spin state is a superstructure along the *c*-axis derived from the $K_2NiF_4$-type order shown in Fig. 2. The canting angle $\eta$ is defined as the deviation from the antiferromagnetic alignment of the magnetic moments for neighboring Ru(1) and Ru(2) ions along the *c*-axis.

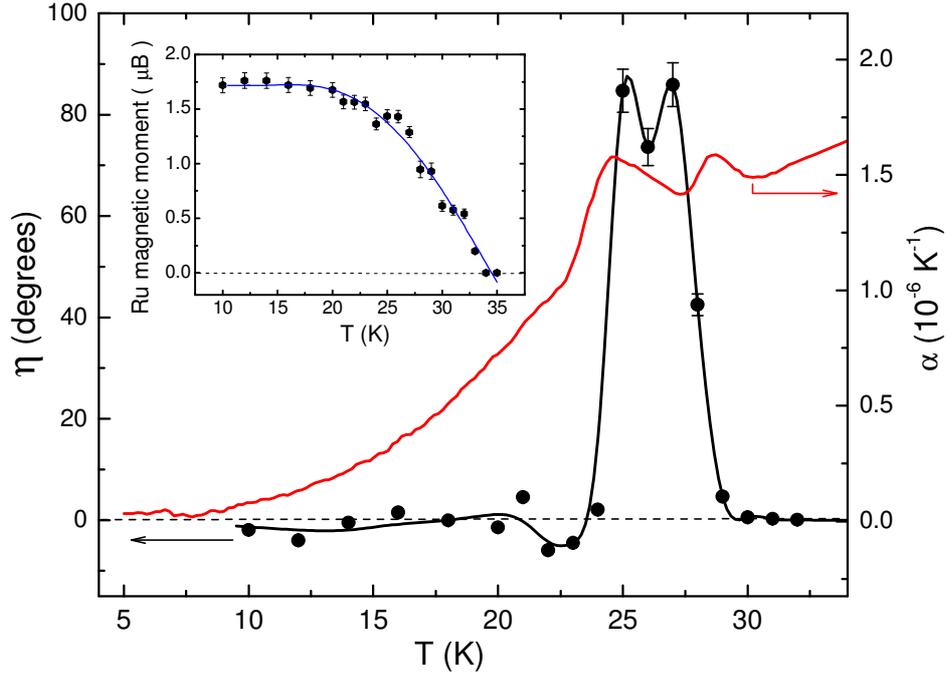

**Fig. 5** Temperature dependence of the canting angle $\eta$ between the magnetic moments of the Ru ions for the non collinear spin state depicted in Fig. 4, and the thermal expansion coefficient, $\alpha(T)$, from Ref. [13] Inset: temperature dependence of the magnitude of the magnetic moment of the Ru ions below the magnetic transition temperature $T_M = 32$ K after refinements of the NPD profiles using the propagation vector $K_{1/2} = (½,½,0)$.

Table I

Fiting parameters of the Rietveld refinements of the $Sr_2YRuO_6$ profiles for selected temperatures using the propagation vectors $K_0 = (0,0,0)$ and $K_{1/2} = (½,½,0)$.

| T (K) | $\chi^2$ | | $R_{wp}$ (%) | | $R_p$ (%) | |
|---|---|---|---|---|---|---|
| | $K_{1/2}$ | $K_0$ | $K_{1/2}$ | $K_0$ | $K_{1/2}$ | $K_0$ |
| 10 | 3.93 | 3.95 | 12.0 | 12.1 | 19.5 | 19.5 |
| 16 | 3.77 | 3.79 | 11.8 | 11.8 | 19.1 | 19.1 |
| 20 | 3.70 | 3.74 | 11.7 | 11.8 | 19.0 | 19.0 |
| 24 | 3.78 | 3.79 | 11.8 | 11.8 | 19.4 | 19.4 |
| 25 | 3.68 | - | 11.6 | - | 19.3 | - |
| 26 | 3.72 | - | 11.7 | - | 19.1 | - |
| 27 | 3.71 | - | 11.8 | - | 19.2 | - |
| 28 | 3.77 | 3.78 | 11.8 | 11.8 | 18.2 | 18.9 |
| 30 | 3.70 | 3.72 | 11.6 | 11.8 | 19.2 | 19.3 |
| 33 | 4.10 | 4.10 | 12.6 | 12.6 | 19.7 | 19.7 |